\documentclass[]{interact}

\usepackage{epstopdf}
\usepackage[caption=false]{subfig}

\usepackage{mathtools}
\usepackage{lineno}
\usepackage{graphics,multirow}
\usepackage{latexsym}
\usepackage{amssymb}     
\usepackage{amsmath}
\usepackage{amsfonts}
\usepackage{enumerate}
\usepackage{lineno}
\usepackage{appendix}
\usepackage{pslatex}
\usepackage{rotating}
\usepackage{tikz}
\usetikzlibrary{matrix,shapes,arrows,positioning,chains}
\usepackage{setspace}
\usepackage[T1]{fontenc}
\usepackage{lmodern}
\usepackage{amsmath}
\usepackage{multirow}
\usepackage{booktabs}
\usepackage{booktabs,caption,fixltx2e}
\usepackage[flushleft]{threeparttable}
\usepackage{wrapfig}
\usepackage{gensymb}
\usepackage{xcolor}

\usepackage[numbers,sort&compress]{natbib}
\bibpunct[, ]{[}{]}{,}{n}{,}{,}

\theoremstyle{plain}

\theoremstyle{definition}

\theoremstyle{remark}

\begin{document}


\title{Penalized Likelihood Approach for the Four-parameter Kappa Distribution}

\author{
\name{Nipada Papukdee\textsuperscript{a},Jeong-Soo Park\textsuperscript{b},and Piyapatr Busababodhin\textsuperscript{ a,*},\thanks{$~^*$ CONTACT Piyapatr Busababodhin, Tel.: +66-43-754244, ORCID: 0000-0003-2056-2804, Email: piyapatr.b@msu.ac.th}}
\affil{\textsuperscript{a}Department of Mathematics, Mahasarakham University, Mahasarakham, Thailand; \\
	\textsuperscript{b}Department of Statistics, Chonnam National University, Gwangju, Korea}
}

\maketitle

\begin{abstract}
The four-parameter kappa distribution (K4D) is a generalized form of some commonly used distributions such as generalized logistic, generalized Pareto, generalized Gumbel, and generalized extreme value (GEV) distributions. Owing to its flexibility, the K4D is widely applied in modeling in several fields such as hydrology and climatic change. For the estimation of the four parameters, the maximum likelihood approach and the method of L-moments are usually employed. The L-moment estimator (LME) method works well for some parameter spaces, with up to a moderate sample size, but it is sometimes not feasible in terms of computing the appropriate estimates. Meanwhile, the maximum likelihood estimator (MLE) is optimal for large samples and applicable to a very wide range of situations, including non-stationary data. However, using the MLE of K4D with small sample sizes shows substantially poor performance in terms of a large variance of the estimator. We therefore propose a maximum penalized likelihood estimation (MPLE) of K4D by adjusting the existing penalty functions that restrict the parameter space. Eighteen combinations of penalties for two shape parameters are considered and compared. The MPLE retains modeling flexibility and large sample optimality while also improving on small sample properties. The properties of the proposed estimator are verified through a Monte Carlo simulation, and an application case is demonstrated taking Thailand's annual maximum temperature data. Based on this study, we suggest using combinations of penalty functions in general.  \footnote{This paper was published at 2022: Papukdee, N., Park, J.-S., \& Busababodhin, P.(2022). Penalized likelihood approach for the four-parameter kappa distribution. Journal of Applied Statistics, 49(6), 1559–1573. https://doi.org/10.1080/02664763.2021.1871592} 
\end{abstract}



\section{Introduction}
\label{sec:intro}

\indent The generalized extreme value (GEV) distribution has been widely used for modeling extremes of natural phenomena, such as hydrological events, and extremes in human society, such as events in insurance and financial markets \citep{coles2001introduction}. In real-life extreme-value applications, samples consist of maxima selected from underlying finite samples and fitting the GEV distribution to the data, which sometimes yields inadequate results. In 1994, Hosking introduced the four-parameter kappa distribution (K4D), a generalization of some common three-parameter distributions such as the GEV, to assist in fitting data when three parameters are not sufficient. K4D has since been used in many fields, particularly in the atmospheric and hydrological sciences (e.g., \cite{HosWal,parida,ParkJung,singh,Wallis,murshed13,ParkSeo,Kjeldsen}).
Moreover, the use of the K4D distribution in simulations where a commitment to a particular three-parameter distribution needs to be avoided may be beneficial (\cite{hosking1993some}). \cite{Blum:2017} found that K4D provides an accurate and reliable representation of daily streamflow across most physiographic regions in the conterminous United States.

\indent The maximum likelihood estimator (MLE) approach and the method of L-moments estimation (LME; \cite{hosking90}) are usually employed for the estimation of the four parameters in K4D. {In general, the LME} shows good performance with small samples and offers some advantages over the MLE, such as easy computation, more robustness, and less bias. The LME has been used for many distributions in various fields \citep{Kjeldsen}. However, using the LME of K4D is sometimes non-feasible in that the estimators cannot be computed within the feasible parameter space \citep{dupuis2001more,ParkJung}. Moreover, the application of the LME is almost limited to stationary data, even though \cite{diebolt2008improving} generalized it to non-stationary cases so as to model temporal covariates. \cite{winchester2000estimation} compared the LME and MLE of K4D by simulation. The MLE is optimal for large samples and can be applied to various situations, including non-stationary data. Nonetheless, we found that the MLE of K4D performs poorly and is unstable {when the sample size is small (n=30) to moderate (n=50)}, especially when one of the scale parameters ($k$) is negative in that the variances of the estimated parameters are considerably large. Such poor performance of the MLE may be due to some intrinsic properties of the K4D or due to the presence of many parameters with constraints, which makes it difficult for the optimization routine to obtain the true solution.

An attempt to solve this problem and to enable a likelihood-based inference led to the development of the MPLE of K4D. Applying the penalty function (PF) to parameters prevents the estimator from having an absurd value. The poor performance of the MLE with a small sample has also been observed in the GEV distribution when one of the shape parameters $(k)$ is negative. For the GEV distribution,
\cite{coles1999likelihood} and \cite{martins2000generalized} proposed PFs on a shape. Their approaches can be applied to two shape parameters in K4D. We modify it for the second shape parameter $(h)$ by adjusting some constants.

\indent This study proposes an MPLE for K4D with PFs on two shape parameters---which are modified from those of \cite{coles1999likelihood} and \cite{martins2000generalized}---and compares it with existing estimation methods. Section 2 describes the K4D. The estimation methods of MLE, LME, and MPLE for K4D are presented in section 3. Section 4 provides a simulation study to compare the performances of the considered estimators. Section 5 illustrates our approach by applying it to the {annual maximum temperature data} in Thailand. Finally, our concluding remarks are presented in section 6. Some details such as technical specifics, tables, and figures are given in the Supplementary Material.

\section{Four parameter kappa distribution} \label{sec:2K4D}

\begin{figure}
	\begin{tabular}{l}	
		\includegraphics[scale = 0.4]{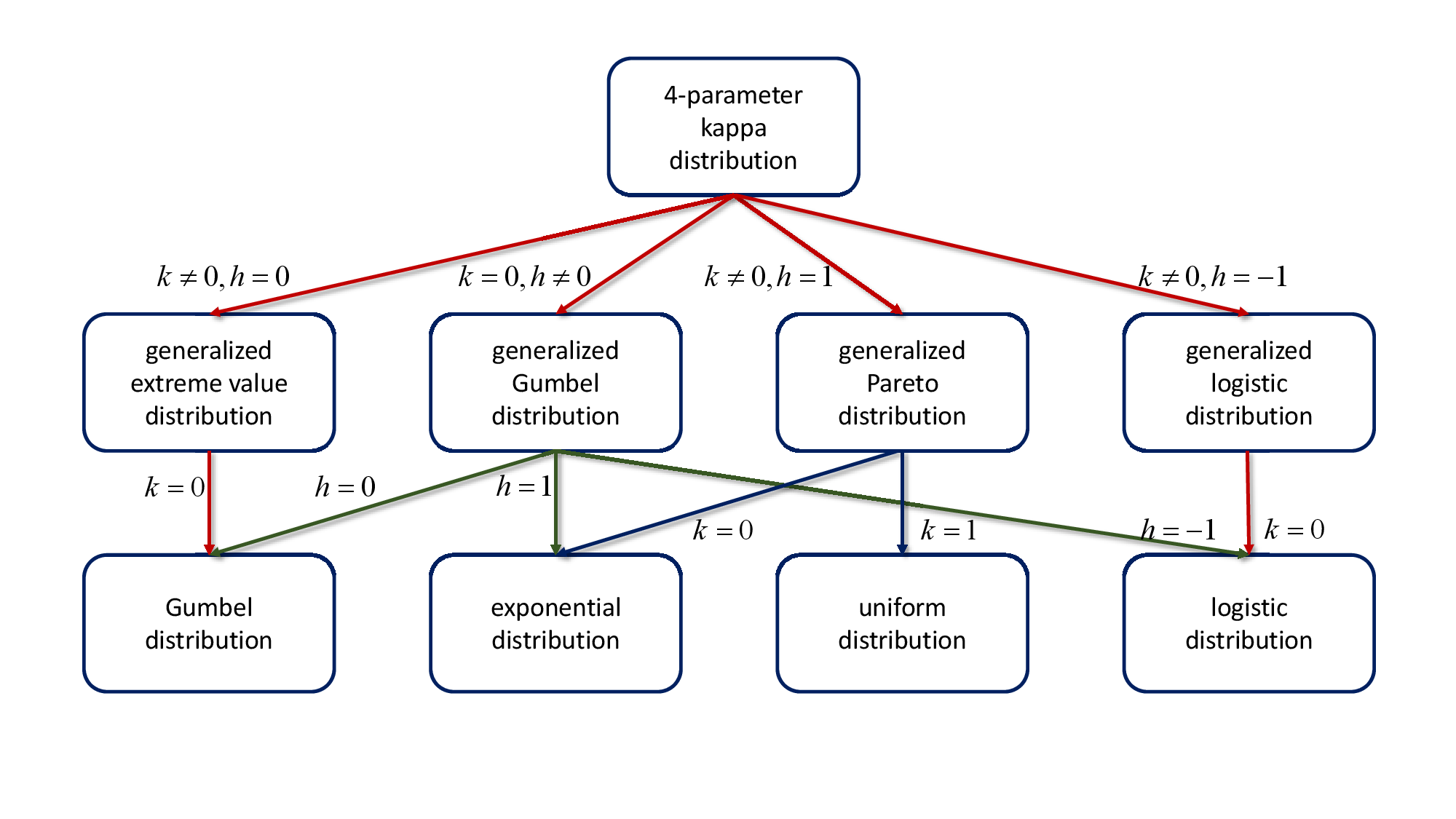}
	\end{tabular}	
	\caption{Relationship of a four-parameter kappa distribution (K4D) to other distributions, which indicates a wide coverage of K4D.}
	\label{K4D_rel}
\end{figure}

The four-parameter kappa distribution, introduced by \cite{hosking1994four}, is a generalized and
reparameterized form of the kappa distribution introduced by
\cite{mkjn73}. K4D includes many distributions as special cases, as shown in
Figure \ref{K4D_rel}: the generalized Pareto distribution for $h=1$, GEVD for
$h=0$, generalized logistic distribution for $h=-1$, and generalized Gumbel distribution for $k=0$, etc.
K4D is therefore very flexible and widely applicable to many situations including not only extreme values but also skewed data.
\cite{parida} and \cite{ParkJung} employed this distribution to analyze annual maximum daily rainfall in India and Korea, respectively. It has also been used for analysis of extreme flood flow series (\cite{murshed13}), and for the maximum wind speed (\cite{moses}). \cite{HosWal}, \cite{Wallis}, and \cite{Kjeldsen} adopted K4D for regional frequency analysis of daily rainfall extremes and flood flow.

The probability density function $(pdf)$ of K4D is, for $\sigma>0,-\infty<x<\infty.$
\[
f(x)=\left\{ \begin{array}{ll}
\sigma^{-1}(1-ky)^{(1/k)-1}(F(x))^{1-h}  & \textrm {if} ~ k\neq0,h\neq0  \\
\sigma^{-1}(1-ky)^{(1/k)-1}F(x)          & \textrm {if} ~ k\neq0,h=0  \\
\sigma^{-1}\textrm {exp}(-y)(F(x))^{1-h} & \textrm {if} ~ k=0,h\neq0  \\
\sigma^{-1}\textrm {exp}(-y)F(x)         & \textrm {if} ~ k=0,h=0,  \\
\end{array} \right.
\]
where
\begin{equation}
F(x) =  {\left\{1-h {[1-k(x-\mu)/\sigma]}^{1/k} \right\}}^{1/h}, \label{CDF1}
\end{equation}
is the cumulative distribution function ($cdf$) and $ y = (x-\mu)/\sigma$.
The quantile function ($x(F)$) of K4D {(inverse cumulative distribution function) as following \cite{hosking1994four}},
\begin{equation}
x(F)= \mu +\dfrac{\sigma}{k} \left\{ 1-\left(\dfrac{1-F^h}{h}\right)^k \right\},\label{XF}
\end{equation}
where $\mu, \sigma$, and $ k,h $ are the location, scale, and two shape parameters, respectively.

\section{Estimation methods}
For the estimation of the four parameters, the maximum likelihood approach and the LME are usually employed; however, it is noteworthy that other methods have also been developed, such as the maximum entropy approach (\cite{singh}) and the so-called LQ-moment technique (\cite{shabri2010lq}).

\subsection{L-moments estimation} \label{sec:sub_lmoments}

The LME was introduced by \cite{hosking90} as a linear combination of expectations of order statistics.
The LME is obtained by solving a system of equations that equates population L-moments to the corresponding sample L-moments. The LME is known to be efficient for parameter estimation of skewed distributions with at least three parameters for small and moderate sample sizes. The LME performs more reliably than the
method of moments estimator and is usually computationally more tractable
than the MLE. The LME works better than the MLE for small sample sizes, and it is less sensitive to outliers (\cite{Hosking85}).
It has been used widely in many fields including in environmental sciences (\cite{Kjeldsen}).

We consider the LME a standard in analyzing skewed data (e.g., extreme
rainfall and flood frequency); the details on the LME are
omitted here. See \cite{hosking1994four} and \cite{HosWal} for the L-moments formula for the K4D.
The LME of K4D has been used for regional frequency analysis in \cite{HosWal} and \cite{Kjeldsen}. \cite{parida} and \cite{moses} used LME of K4D for modeling summer monsoon rainfall in India and monthly maximum wind speed in Botswana, respectively.
\cite{murshed13} considered the higher-order L-moment (LH) estimator of K4D, applying it to annual maximum flood and sea level data.
\cite{asquith2007moments} compared K4D with other four-parameter asymmetric distributions with L-moments.
In calculating the LME of K4D, we use the R package `lmom' developed by \cite{hosking90}.

One disadvantage of the LME is that Newton-Raphson type algorithms to solve systems of L-moments equations sometimes fail to converge. The percentage of failure has been reported to be up to 40\% (\cite{dupuis2001more}). We observe 12 failures in the datasets of the annual maximum daily rainfall from 25 weather stations in northeastern Thailand. In these cases, the MPLEs are a reasonable alternative.

\subsection{Maximum likelihood estimation}

Assuming observations $ (X_1,X_2,...,X_n) $ follow the K4D, the negative log likelihood function is
\begin{equation}
-l(\mu,\sigma,k,h) =  n\ln\sigma-(1-h)\sum_{i=1}^n\ln F(x_i)-\dfrac{1-k}{k}\sum_{i=1}^n\ln G_i,  \label{NLH}
\end{equation}
under $ k\neq0 $ and $ h\neq0 $, where $ G_i = 1-{k(x_i-\mu)}/{\sigma}$ for non-zero $\sigma$.
The MLEs are obtained by minimizing (\ref{NLH}) with respect to $\mu,~\sigma,~k,~h$. Since no explicit minimizer is available,
this function is iteratively minimized using a Newton-type optimization algorithm.
In the numerical iterations for the MLE, $ G_i $ and $F(x_i)$ go below zero sometimes, and the routine thus fails to obtain a solution. To overcome this difficulty in the optimization algorithm, \cite{park2001modelling}  assigned a penalty for the infeasible region of parameters.
The K4D was fitted to annual maximum daily precipitation by the MLE in \cite{ParkJung}. \cite{ParkKim} provided the Fisher information matrix for K4D.
\cite{dupuis2001more} compared the performance of MLE and LME for the K4D.
\section{Maximum penalized likelihood estimation}

For small sample sizes, the MLE of the GEV distribution sometimes underestimates the negative value of the shape parameter ($ k $). Consequently, it causes a large bias and variance in
extreme upper quantiles. To overcome this problem and for a likelihood-based inference, \cite{coles1999likelihood} and \cite{martins2000generalized} proposed the MPLE, which assigns a penalty for a large negative value of $ k $.
To obtain the MPLE in GEVD, the penalized negative log-likelihood to be minimized is
\begin{equation} \label{pnllk_gev}
l_{pen}(\mu,\sigma,k)=  -ln (L(\mu,\sigma,k))- ln ( p(k)) ,
\end{equation}
where $p(k)$ is a penalty function (PF) on $k$.

\begin{figure}
	\begin{center}
		\includegraphics[scale=0.65]{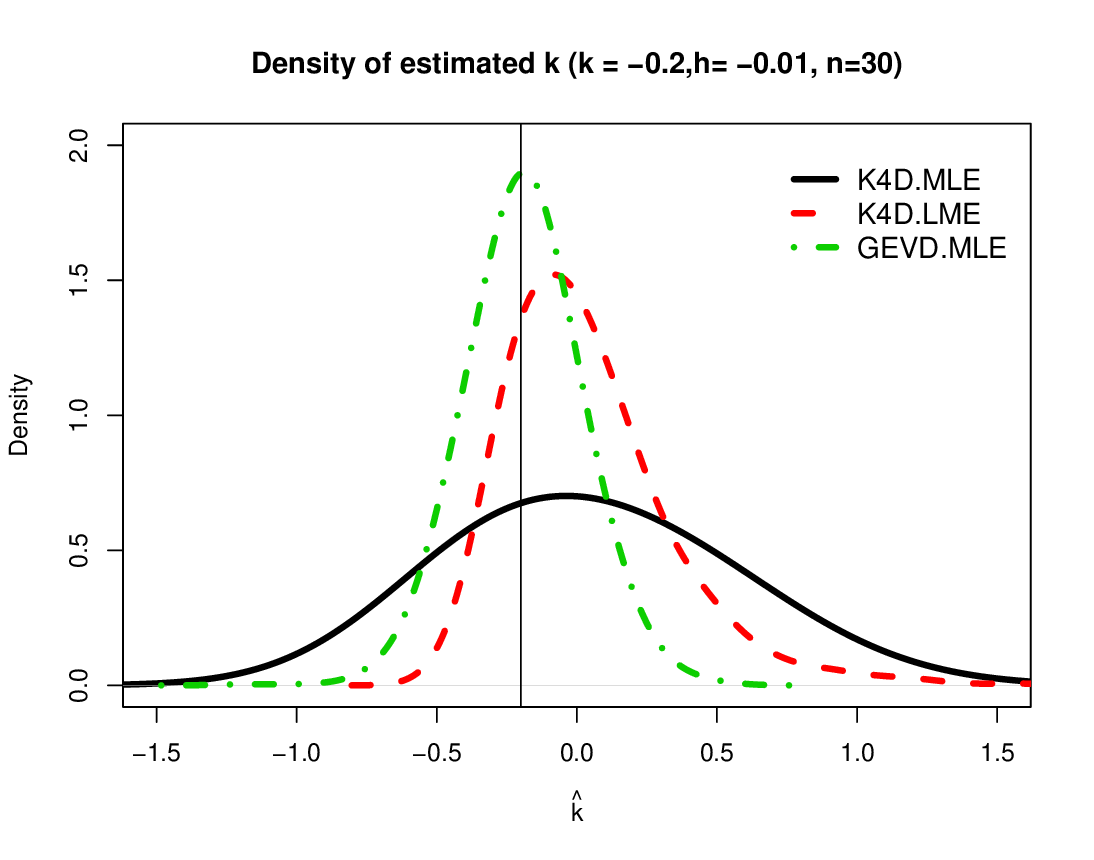}\\
		\caption{Kernel density plot of the sampling distribution of $\hat{k}$ which were estimated using three different methods with sample size $n=30$. The vertical line represents the true value of $k=-0.2$ when $h=-0.01$.}
		\label{density}
	\end{center}
\end{figure}

Since the K4D is a generalization of the GEVD, we focus on the poor performance of the MLE of the shape parameters in K4D. In fact, we find that the MLE of parameter $ k $ has a relatively large variance compared to the LME. Figure \ref{density} shows the kernel density estimates of the sampling distribution of $\hat {k}$ when the true value of $ k $ is $ -0.2 $ for a fixed $ h=-0.01 $.
The LME sometimes fails to obtain the estimates due to convergence failure. We therefore attempt to address this difficulty by employing a maximum penalized likelihood approach for K4D.

\subsection{Penalty for $k$}
\cite{coles1999likelihood} proposed the following exponential PF in the GEVD:
\begin{equation}\label{p_coles}
p(k) = \begin{cases} 1 & \mbox{if $k \geq 0$}\\
\exp\{-\lambda(\frac{1}{1+k}-1)^{\alpha}\} & \mbox{if $-1< k < 0$}\\
0 & \mbox{if $k\leq -1$}\end{cases}
\end{equation}
for non-negative values of $\alpha$ and $\lambda$. For the hyperparameters $\alpha$ and $\lambda$,
they suggested using a combination of $\alpha=1$ and $\lambda=1$.
We call this function CD.

\cite{martins2000generalized} proposed the following penalty, a
Beta probability density function (pdf) on $k$ between $-0.5$ and $0.5$:
\begin{eqnarray} \label{MSpenalty}
p(k)=(0.5+k)^{p-1} (0.5-k)^{q-1} /B(p,q),
\end{eqnarray}
where $B(p,q)$ is the beta function. They chose $p=6$ and $q=9$ based on their prior
hydrological information and experiments. We abbreviate this function as MS. This is considered a prior in Bayesian analysis. MPLE therefore corresponds to the mode of the Bayesian posterior distribution of the parameter. Meanwhile, \textcolor{red}{\cite{park2005simulation}} suggested using $p=2.5$ and $q=2.5$.

\subsection{Penalty for $h$}
For the penalty for $ h $,
{we} change constants from CD and MS to adjust the range of $ h \in (-1.2,\; 1.2)$.
This range is adopted from \cite{dupuis2001more}.
Coles and Dixon's adjusted PF for $ h $ is
\begin{eqnarray}
p(h) & = &\left\{\begin{array}{ll}
1  & \textrm {if} ~ h \geq 0 \\
\textrm {exp}\left\{ -\lambda\left(\dfrac{1}{1.5+h}-0.67\right)^{\alpha} \right\}&
\textrm {if} ~ -1.2 < h <0,  \\
0 & \textrm {if} ~ h \leq -1.2   \label{pencd}
\end{array} \right.
\end{eqnarray}

Martines and Stedinger's adjusted penalty for $ h \in (-1.2,\; 1.2)$ is
\begin{eqnarray}
p(h) & = &  \dfrac{(1.2+h)^{p-1}(1.2-h)^{q-1}}{B_E (p,q)}, \label{penms}
\end{eqnarray}
with $p=6$ and $q=9$, where
\begin{eqnarray*}
	B_E (p,q) &=& \int_{-1.2}^{1.2} (1.2+h)^{p-1}(1.2-h)^{q-1} dh.
\end{eqnarray*}

\begin{figure}[h]	
	\centering
	\hspace{0.cm}
	\includegraphics[scale=0.6]{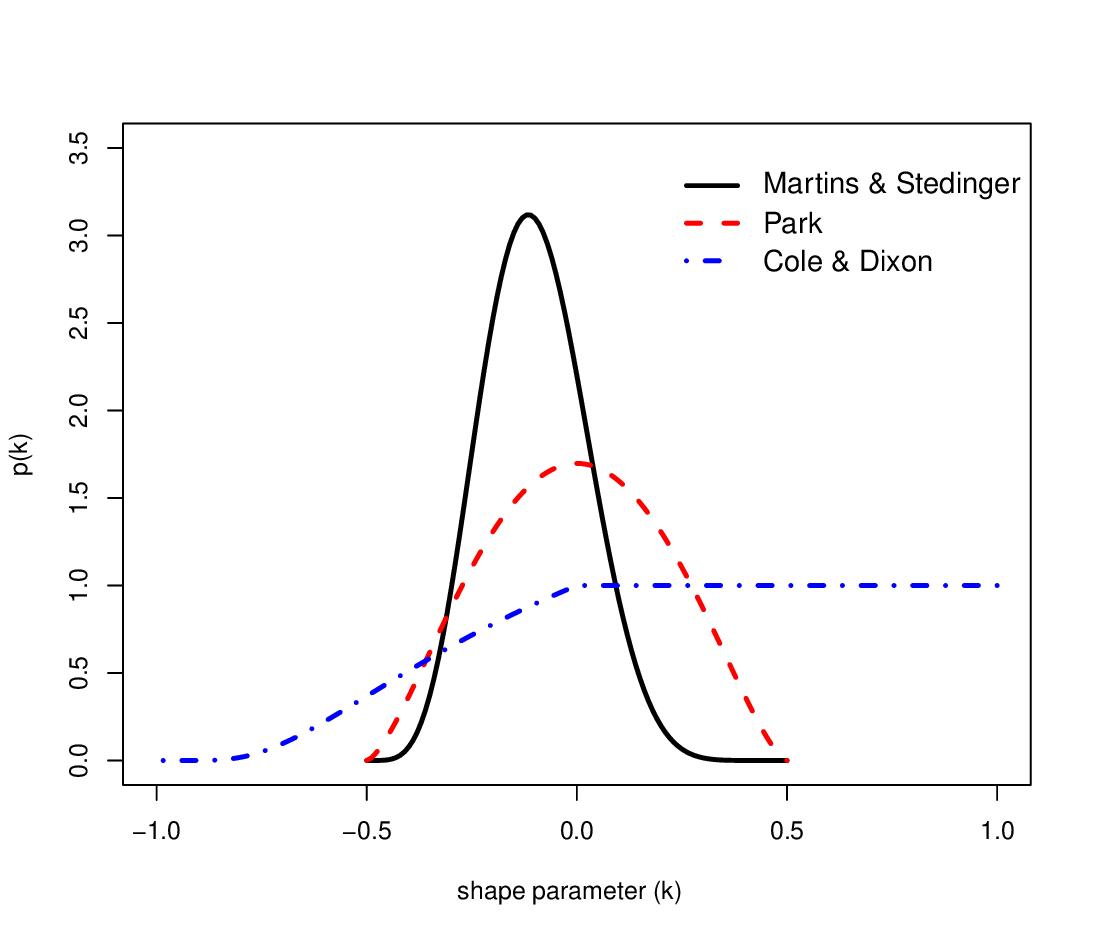}
	\hspace{0.cm}
	\caption{Three different penalty functions for shape parameter $k$.}
	\label{penaltypk}
\end{figure}

\begin{figure}[h]	
	\centering
	\hspace{0.cm}
	\includegraphics[scale=0.6]{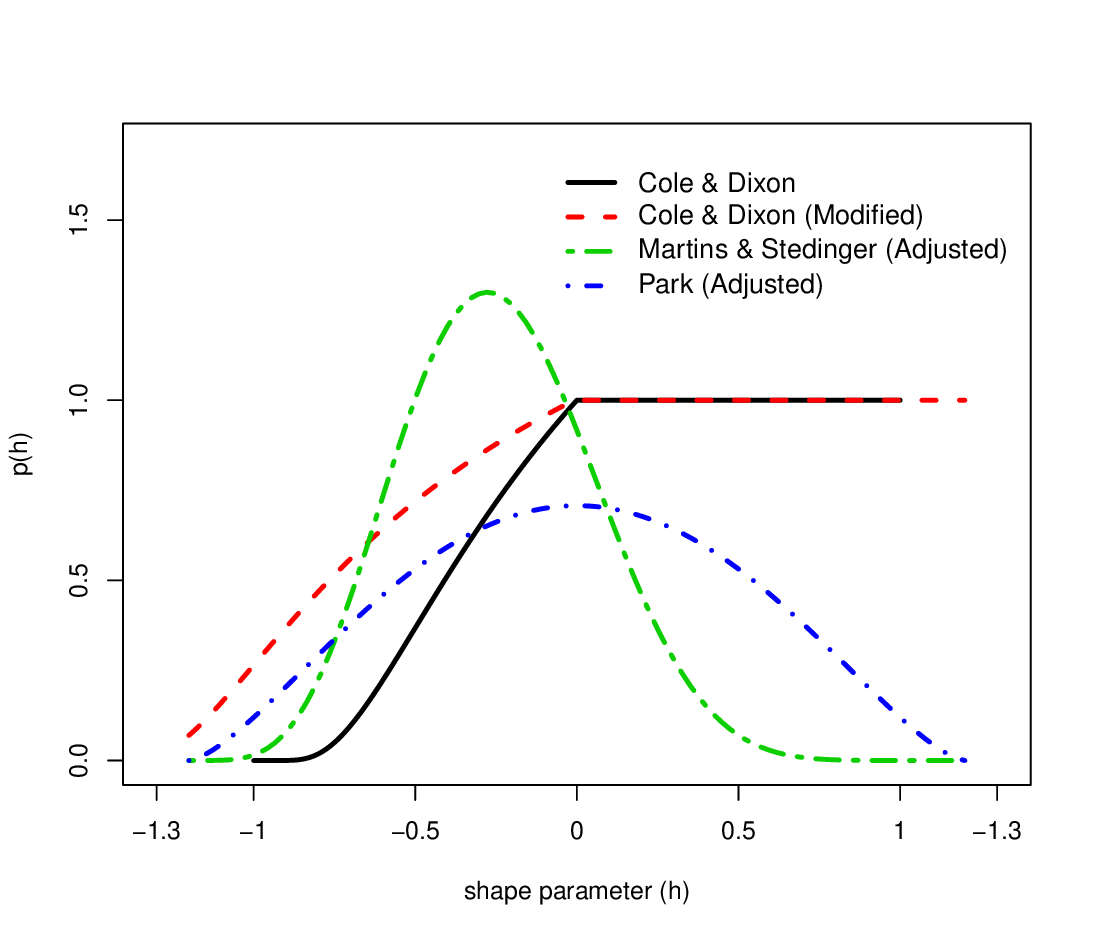}
	\hspace{0.cm}
	\caption{Four different penalty functions for shape parameter $ h $. }
	\label{penaltyph}
\end{figure}

\noindent Park's adjusted penalty is the same, but with $p=2.5$ and $q=2.5$.
Figures \ref{penaltypk} and \ref{penaltyph} show the graphs of the PFs plotted against $ k $ and $ h $, respectively.

We differentiate these penalties from the original ones by terming them `adjusted penalties'.
Note that the original penalties are defined on the range $(-.5, .5)$, whereas the adjusted penalties are
on $(-1.2, 1.2)$. These penalties are denoted by CD$_o$(h), ~MS$ _o$(h),~ P$ _o $(h), and ~CD$ _a$(h),~ MS$ _a$(h), ~ P$ _a$(h), where the subscripts "$o$" and "$a$" denote "original" and "adjusted," respectively. We now have six penalties for $h$.

\subsection{Joint penalty for $k$ and $h$}
The penalized negative log likelihood function for K4D is
\begin{equation}
l_{pen}(\mu,\sigma,k,h) = - \textrm{ln}(L(\mu,\sigma,k, h)) - \textrm{ln}(p(k,h)). \label{MPLE1}
\end{equation}
where $ p(k,h) $ is a joint PF of $ k $ and $ h $. In this study, we treat $ k $ and $h$ as independent, so that $ p(k,h) = p(k)\times p(h) $. The other types of joint PFs can be considered in future research.

This study considers combinations of three penalties for $k$ and six penalties for $h$. Thus, 18 $(3 \times 6)$ penalties are employed. We denote MPLE.P$_o$(k)CD$ _a$(h), for example, which symbolizes the MPLE with Park's penalty for $k$ and the adjusted CD penalty for $h$.


\section{Monte Carlo Simulation}
Monte Carlo simulations are performed to investigate the effect of using the MLE, LME, and MPLE on the estimation of extreme quantiles: x(F=0.90), x(F=0.95), x(F=0.99), x(F=0.995), and x(F=0.999) for sample sizes of 30 and 50. A total of 1,000 repetitions are executed to obtain the relative bias (RBIAS) and relative root mean square error (RRMSE) of the quantile estimators. The RBIAS and RRMSE are defined by \cite{castillo2005extreme} as
\begin{equation}
\textrm {RBIAS} = \dfrac{1}{M}\sum_{j=1}^M \left(\dfrac{q_{j}^{e}-q^{t}_{j}}{q^{t}_{j}} \right)~~  \textrm{and} ~~  \textrm {RRMSE} = \sqrt{\dfrac{1}{M}\sum_{j=1}^M \left(\dfrac{q_{j}^{e}-q^{t}_{j}}{q^{t}_{j}}\right)^2},
\end{equation}
where $ M $ is the number of successful convergences among 1,000 trails, and $ q_{j}^{e} $ and $ q_{j}^{t} $ are the estimated and true quantiles, respectively. In interpreting the result, we focus more on RRMSE than on RBIAS.

The simulations are performed by generating random numbers from the K4D, with $ \mu $ and $ \sigma $ fixed at $ 0 $ and $ 1 $, respectively. Values of $ k \in [-0.4,0.4] $ and $ h \in (-1.2,1.2) $ are considered for sample sizes $n=30$ and $n=50$. The RBIAS and RRMSE are calculated for the estimated quantiles of x(F=0.90),~ x(F=0.95),~ x(F=0.99),~ x(F=0.995), and x(F=0.999), which were computed by the MLE, LME, and 18 types of MPLE methods.

\begin{table}[h!]
	\small
	\centering
	\scalebox{1.3}{
		\begin{threeparttable}
			\caption{The relative root mean squared errors of the quantile $x(F)$ estimates calculated from the MLE, LME, and four best MPLE methods where data are generated from the four-parameter kappa distribution with $k=-0.2$ and $h=-0.2$ and sample size $n=30$. MLE = maximum likelihood estimations, LME = L-moments estimations, and MPLE = maximum penalized likelihood estimation.}
			\label{tb1:rrmse1_n30}
			\begin{tabular}{@{}lrrrrr@{}}
				\toprule
				\multicolumn{1}{c}{\multirow{2}{*}{~~Methods~~~~~}} & \multicolumn{5}{c}{$ x(F) $}  \\
				\cmidrule(l){2-6}
				\multicolumn{1}{c}{} &
				\begin{tabular}[c]{@{}l@{}}~~0.90~~ \end{tabular} & \begin{tabular}[c]{@{}l@{}}~~~0.95~~ \end{tabular} & \begin{tabular}[c]{@{}l@{}}~~0.99~~ \end{tabular} & ~0.995~~ & ~0.999~~  \\
				\midrule
				MLE  							& ~0.0783	&  ~0.0977		& ~0.2289	& ~0.2857	& ~1.7546\\
				LME							& ~0.0736	&  ~0.0836		& ~0.1611	& ~0.1895 	& ~0.3242\\
				MPLE.MS$ _o$(k)MS$ _o$(h) 	    & 0.0498	&  0.0512		& 0.0658	& 0.0766 	& ~0.1021\\
				MPLE.MS$ _o$(k)MS$ _a$(h) 	    & ~0.0512	&  ~0.0532		& ~0.0676	& ~0.0773	& 0.1005\\
				MPLE.MS$ _o $(k)P$ _o $(h) 	& ~0.0509	&  ~0.0514		& 0.0658	& ~0.0769 	&~0.1048\\
				MPLE.MS$ _o $(k)P$ _a $(h) 	& ~0.0620	&  ~0.0618		& ~0.0720	& ~0.0791	&~0.1068\\
				\noalign{\smallskip}\hline
			\end{tabular}
	\end{threeparttable}}
\end{table}

\begin{table}[h!]
	\small
	\centering
	\scalebox{1.3}{
		\begin{threeparttable}
			\caption{Same as Table \ref{tb1:rrmse1_n30} but $h=0.2$.}
			\label{tb2:rrmse2_n30}
			\begin{tabular}{@{}lrrrrr@{}}
				\toprule
				\multicolumn{1}{c}{\multirow{2}{*}{~~Methods~~~~~}} & \multicolumn{5}{c}{$ x(F) $}  \\
				\cmidrule(l){2-6}
				\multicolumn{1}{c}{} &
				\begin{tabular}[c]{@{}l@{}}~~0.90~~ \end{tabular} & \begin{tabular}[c]{@{}l@{}}~~~0.95~~ \end{tabular} & \begin{tabular}[c]{@{}l@{}}~~0.99~~ \end{tabular} & ~0.995~~ & ~0.999~~  \\
				\midrule
				MLE  							& 0.0787 & 0.1132 & 0.6489 & 0.4533 & 1.0818  \\
				LME							& 0.0778 & 0.0803 & 0.1586 & 0.1931 & 0.3889  \\
				MPLE.MS$ _o$(k)MS$ _o$(h) 	& 0.0536 & 0.0554 & 0.0720 & 0.0776 & 0.0949  \\
				MPLE.MS$ _o$(k)MS$ _a$(h) 	& 0.0543 & 0.0563 & 0.0729 & 0.0782 & 0.0945  \\
				MPLE.MS$ _o $(k)P$ _o $(h) 	& 0.0523 & 0.0525 & 0.0702 & 0.0768 & 0.0981  \\
				MPLE.MS$ _o $(k)P$ _a $(h) 	& 0.0629 & 0.0594 & 0.0734 & 0.0780 & 0.1012   \\
				\noalign{\smallskip}\hline
			\end{tabular}
	\end{threeparttable}}
\end{table}

\newpage
\begin{table}[!htbp]
	\small
	\centering
	\scalebox{1.3}{
		\begin{threeparttable}
			\caption{Same as Table \ref{tb1:rrmse1_n30} but $k=0.4$ and $h=-0.5$.}
			\label{tb3:rrmse3_n30}
			\begin{tabular}{@{}lrrrrr@{}}
				\toprule
				\multicolumn{1}{c}{\multirow{2}{*}{~~Methods~~~~~}} & \multicolumn{5}{c}{$ x(F) $}  \\
				\cmidrule(l){2-6}
				\multicolumn{1}{c}{} &
				\begin{tabular}[c]{@{}l@{}}~~0.90~~ \end{tabular} & \begin{tabular}[c]{@{}l@{}}~~~0.95~~ \end{tabular} & \begin{tabular}[c]{@{}l@{}}~~0.99~~ \end{tabular} & ~0.995~~ & ~0.999~~  \\
				\midrule
				MLE  							& ~0.0219	&  ~0.0159		& ~0.0255	& ~0.0335	&~0.0459\\
				LME							& ~0.0194	&  ~0.0164		& ~0.0257	& ~0.0316 	&~0.0446\\
				MPLE.CD$ _o $(k)P$ _o $(h) 	& ~0.0182	&  0.0142		& ~0.0251	& ~0.0321	&~0.0394\\
				MPLE.P$ _o$(k)P$ _o$(h)  		& 0.0179	&  ~0.0143		& ~0.0190	& ~0.0227	&~0.0333\\
				MPLE.P$ _o $(k)CD$ _o $(h) 	& ~0.0181	&  ~0.0143		& ~0.0187	& 0.0222	&~0.0320\\
				MPLE.P$ _o $(k)MS$ _a $(h) 	& ~0.0184	&  ~0.0143		& ~0.0193 	& ~0.0243	&~0.0410\\
				\noalign{\smallskip}\hline
			\end{tabular}
	\end{threeparttable}}
\end{table}

\begin{table}[!htbp]
	\small
	\centering
	\scalebox{1.3}{
		\begin{threeparttable}
			\caption{Same as Table \ref{tb1:rrmse1_n30} but $k=0.4$ and $h=0.5$.}
			\label{tb4:rrmse4_n30}
			\begin{tabular}{@{}lrrrrr@{}}
				\toprule
				\multicolumn{1}{c}{\multirow{2}{*}{~~Methods~~~~~}} & \multicolumn{5}{c}{$ x(F) $}  \\
				\cmidrule(l){2-6}
				\multicolumn{1}{c}{} &
				\begin{tabular}[c]{@{}l@{}}~~0.90~~ \end{tabular} & \begin{tabular}[c]{@{}l@{}}~~~0.95~~ \end{tabular} & \begin{tabular}[c]{@{}l@{}}~~0.99~~ \end{tabular} & ~0.995~~ & ~0.999~~  \\
				\midrule
				MLE  						& ~0.0156 & ~0.0139 & ~0.0237 & ~0.0312 & ~0.0758 \\
				LME							& ~0.0153 & ~0.0135 & ~0.0253 & ~0.0353 & ~0.0664\\
				MPLE.CD$ _o$(k)CD$ _a$(h) 	& ~0.0146 & ~0.0127 & ~0.0217 & ~0.0264 & ~0.0369\\
				MPLE.CD$ _o $(k)P$ _a $(h) 	& ~0.0146 & ~0.0132 & ~0.0212 & ~0.0264 & ~0.0357\\
				MPLE.P$ _o$(k)P$ _a$(h) 	& ~0.0150 & ~0.0127 & ~0.0220 & ~0.0307 & ~0.0527\\
				MPLE.P$ _o $(k)CD$ _o $(h) 	& ~0.0146 & ~0.0133 & ~0.0228 & ~0.0317 & ~0.0463\\
				\noalign{\smallskip}\hline
			\end{tabular}	
	\end{threeparttable}}
\end{table}

\begin{figure}
	\begin{center}
		\includegraphics[scale=0.6]{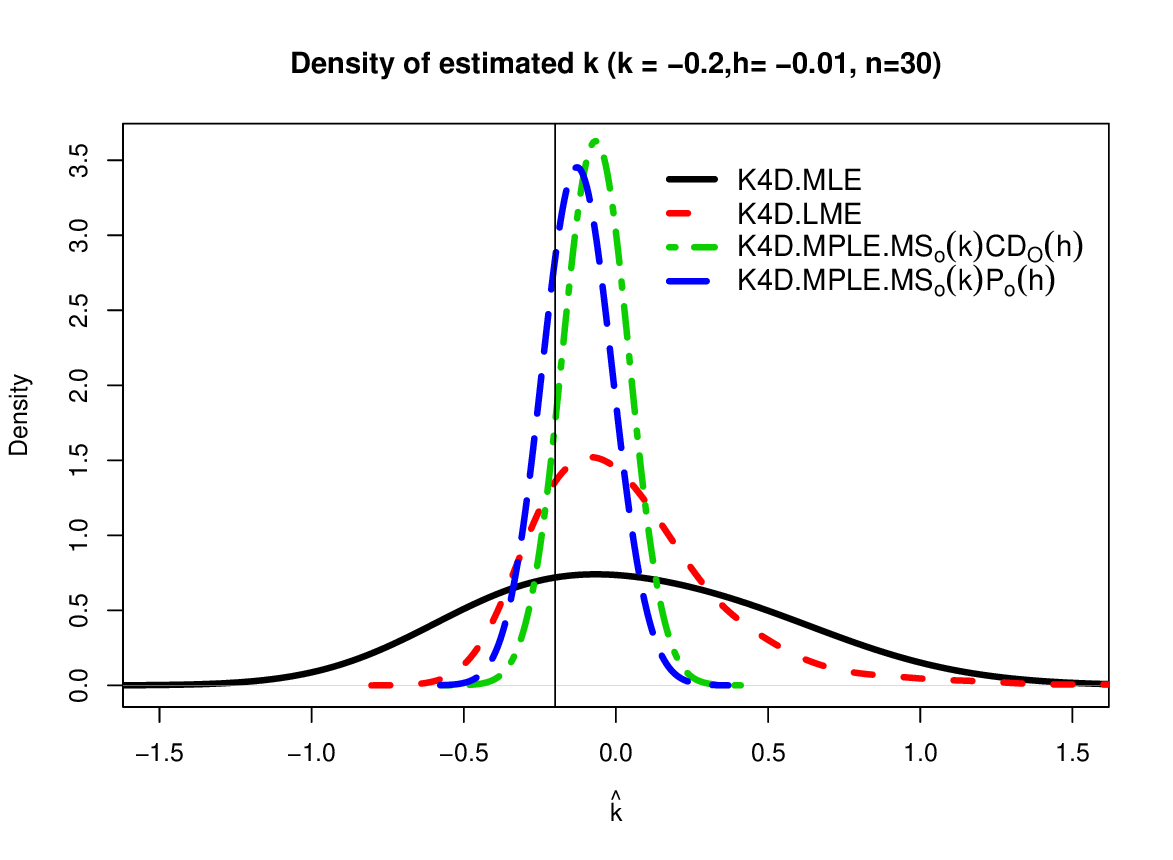}\\
		\caption{Same as Figure \ref{density} but two maximum penalized likelihood estimation (MPLE) methods.}
		\label{kernal1}
	\end{center}
\end{figure}

\begin{figure}
	\begin{center}
		\begin{tabular}{c}
			\includegraphics[width=8.5cm, height=7.8cm]{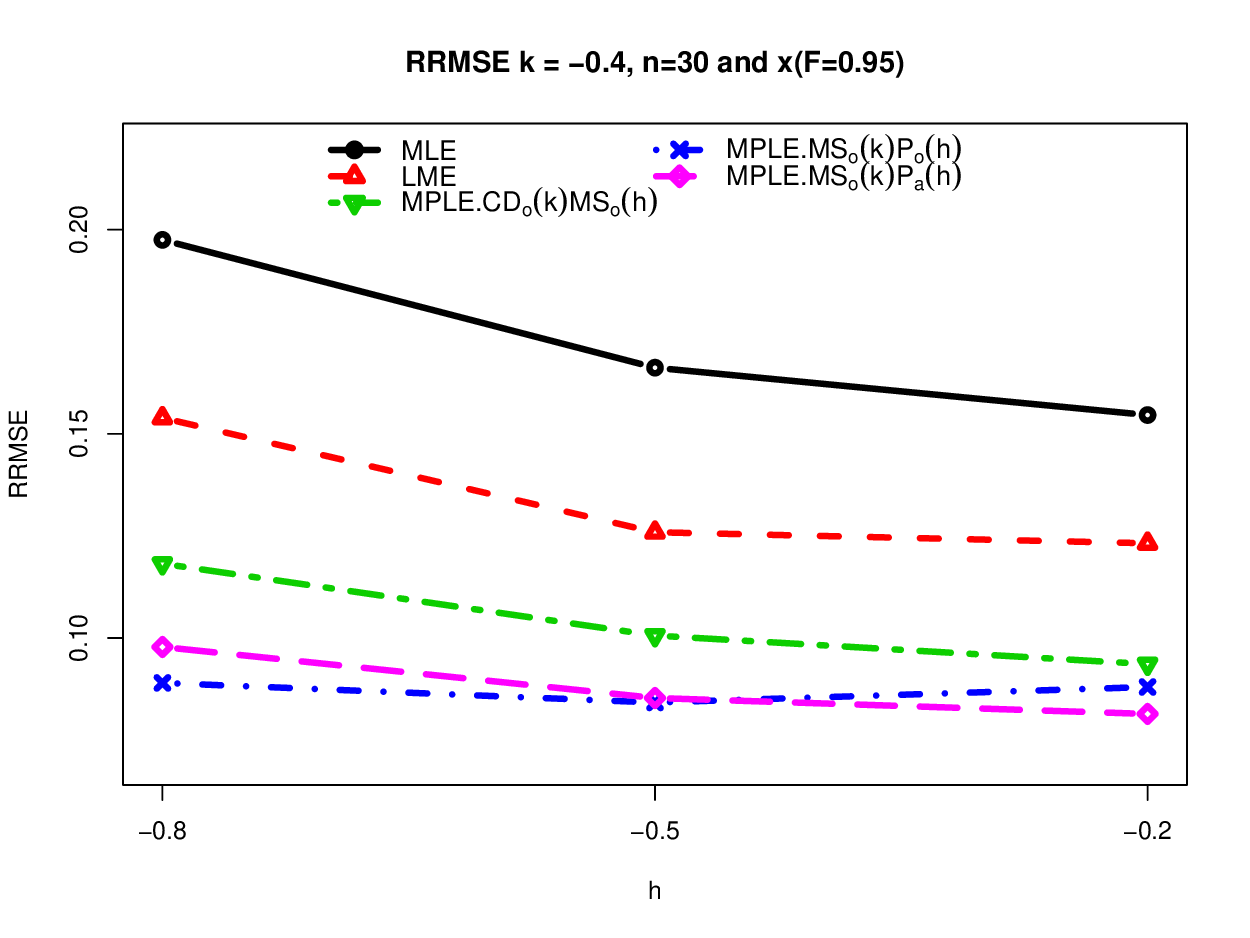}
			\includegraphics[width=8.5cm, height=7.8cm]{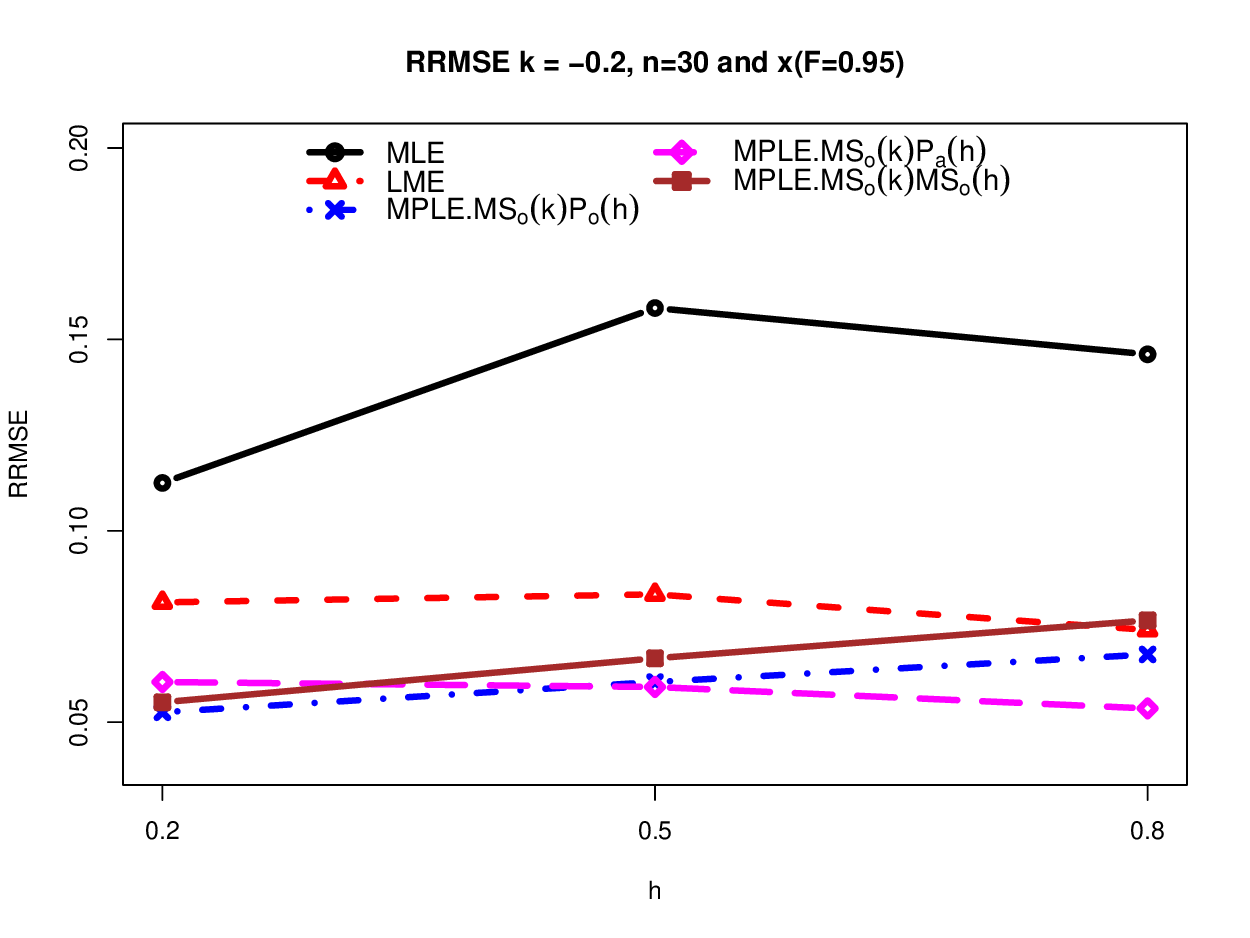}\\
			\includegraphics[width=8.5cm, height=7.8cm]{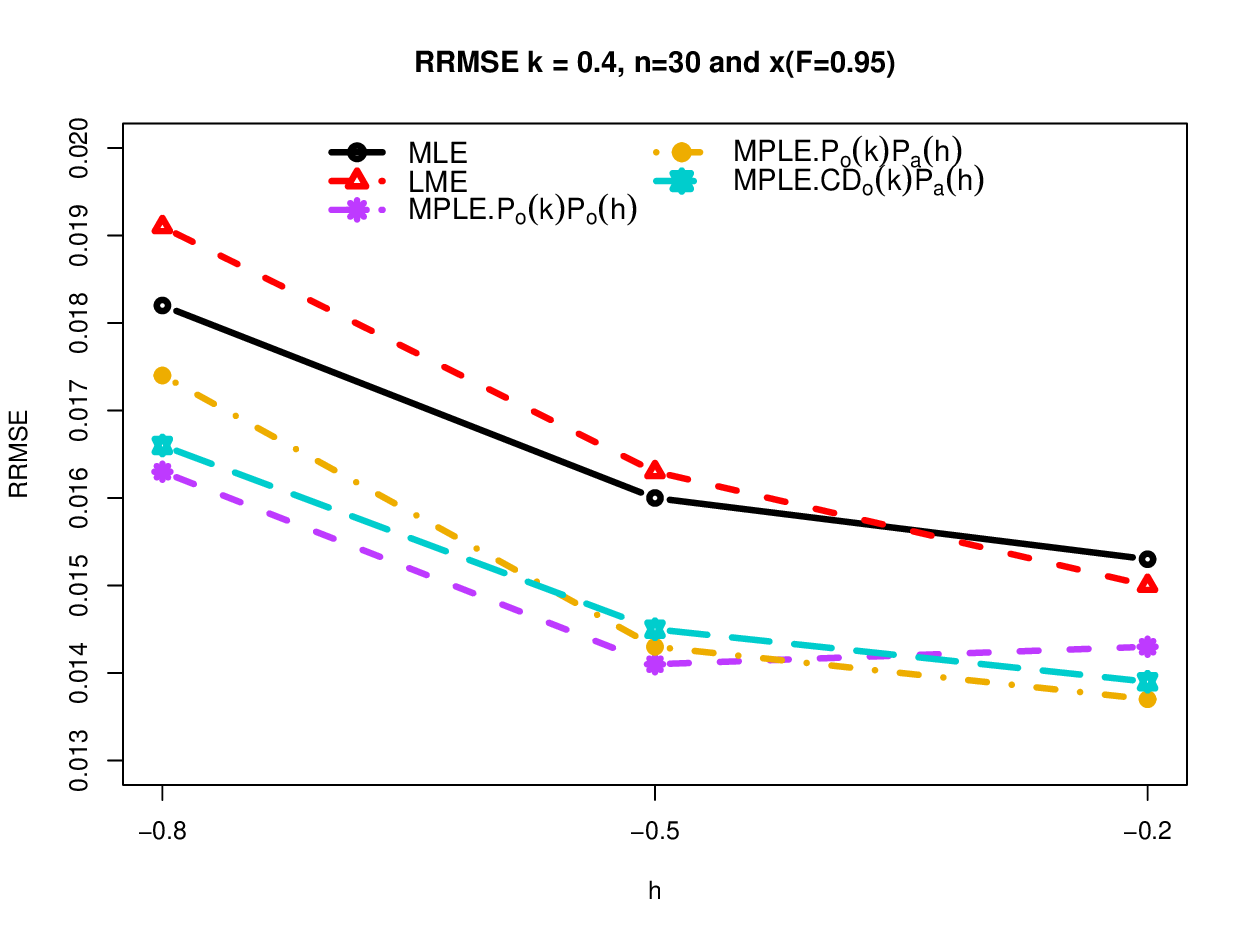}
			\includegraphics[width=8.5cm, height=7.8cm]{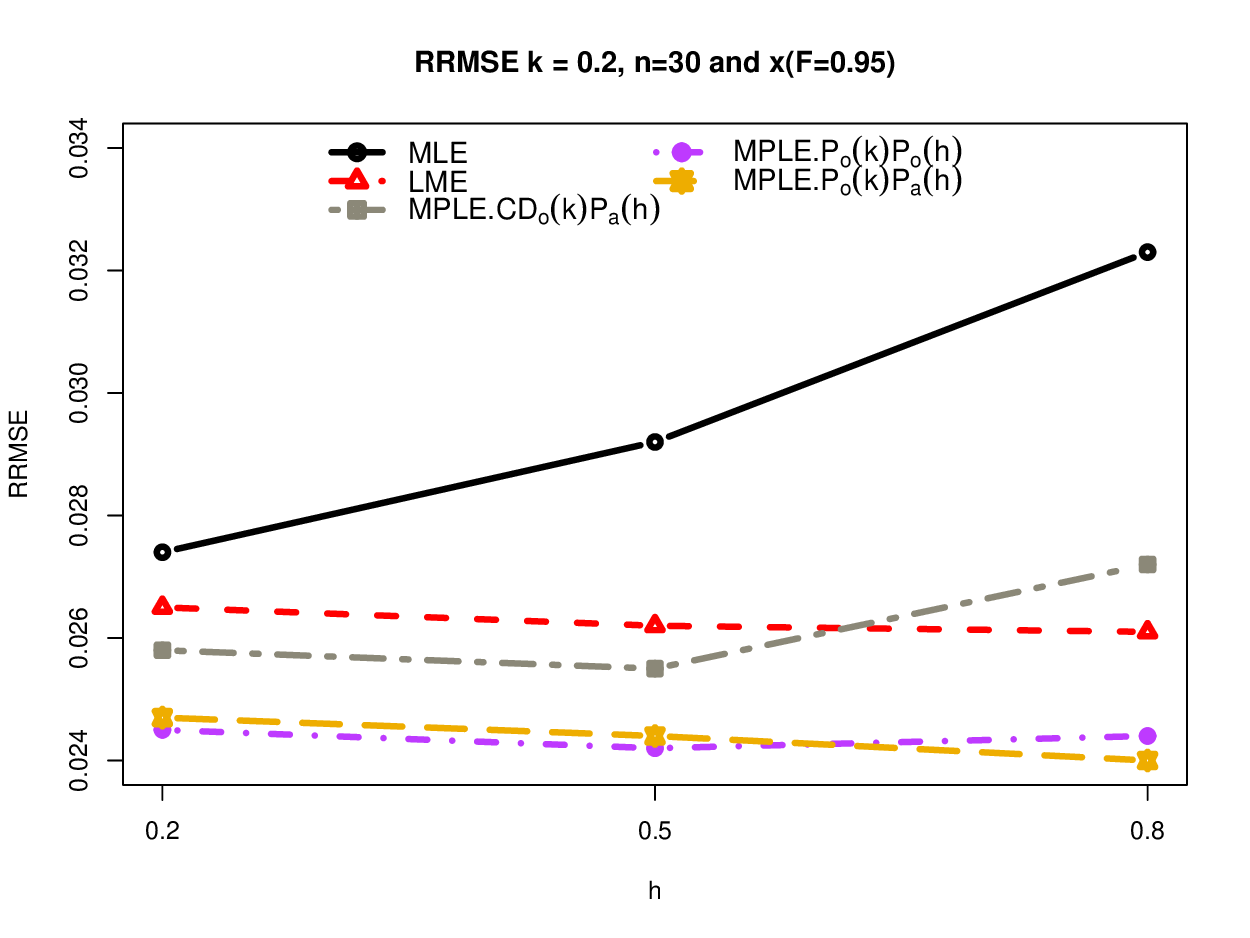}
		\end{tabular}
		\caption{Same as Table \ref{tb1:rrmse1_n30} but for 0.95 quantile estimates for various values of $k$ and $h$.}
		\label{RRMSE_095_n30_fig5}
	\end{center}	
\end{figure}

\indent Tables \ref{tb1:rrmse1_n30} to \ref{tb4:rrmse4_n30} report the RRMSE values of the estimation of the four parameters obtained by the six estimation methods for $(k=-0.2, h=-0.2)$, $(k=-0.2, h=0.2)$, $(k=0.4, h=-0.5)$, and $(k=0.4, h=0.5)$ when $n=30$. Figure \ref{RRMSE_095_n30_fig5} shows the RRMSE of the 0.95 quantile estimators of K4D for sample size $n=30$ for the estimation methods of MLE, LME, and the best three MPLEs. Figure \ref{kernal1} shows the kernel density estimates of the sampling distribution of $\hat{k}$, which are estimated by four different methods. Thus, we now see the MPLEs work well in estimating $k$.

\section{Application to the annual maximum temperature at Surin, Thailand}\label{sec:6HydroData}
The annual maximum temperature data for Surin in Thailand are taken, recorded over the period of 1990 to 2018. The estimation methods considered in this study are employed for K4D fit.
The Anderson-Darling (AD) and Kolmogorov-Smirnov (KS) statistics and a modified prediction absolute error (MPAE) are used to assess the goodness-of-fit of the methods.
The MPAE is defined as follows:
\begin{equation}\label{mpae}
\text{MPAE}= \frac{1}{n}\sum^{n}_{i = 1}|x_{(i)}-\hat{x}_{(i)}|,
\end{equation}
where $n$ is the number of observations, $x_{(i)}$ are the ascending-order observations, and $\hat{x}_{(i)}$ are the estimated quantiles obtained from Eq (\ref{XF}) under the plotting position $p_{i} = (i-0.35)/n$ \citep{HosWal}. The MPAE is modified from the expected prediction squared error of {\cite{efron1994introduction}}.

\begin{table}[!htbp]
	\caption{Estimates of parameters, standard errors in parentheses, goodness-of-fit statistics, and p-values in parentheses, in which the four-parameter kappa distribution was fitted to the annual maximum daily temperature in Surin by the MLE, LME, and four best MPLE methods. MLE = maximum likelihood estimations, LME = L-moments estimations, and MPLE = maximum penalized likelihood estimation. MAPE = modified prediction absolute error, AD = Anderson-Darling Statistic, and KS = Kolmogorov-Smirnov Statistic.}\vspace{0.3cm}
	\label{temperature}
	\scalebox{.9}{
		\begin{tabular}{lccccccc}
			\hline\noalign{\smallskip}
			\textbf{Methods }& \hspace{0.1cm}$\hat\mu$ & \hspace{0.1cm} $\hat\sigma$ & \hspace{0.1cm}$\hat k$ &\hspace{0.1cm} $\hat h$ &  MPAE &\hspace{0.1cm} $AD$ & \hspace{0.1cm}$KS$ \\ [0.1ex]
			\noalign{\smallskip}\hline\noalign{\smallskip}
			MLE       	&  39.457  & ~0.3550  &  -0.2436  & -4.3755   & 0.4077  	& 1.7206   & 0.2200   \\
			& (0.024)  & (0.0284) &  (0.0140) & (0.0774)  &             & (0.1319) & (0.1208) \\
			
			LME        	& 39.222   &  0.6915 &  -0.2552  	&  -0.4391 	& 0.7519   &  0.9576  &  0.1879  \\
			& (0.009)  & (0.0023) &  (0.0063)  	& (0.0042)  &          & (0.3795) & (0.2575)  \\

			MPLE.CD$ _o $(k)MS$ _a $(h)  &  39.040  & 0.6409  	&  -0.0309   	& -0.2232  	& 0.2369 	&  0.3248  &  0.1091  \\
			& (0.006)  & (0.0176)  &  (0.0074)  	& (0.0041)  &           & (0.9178 )& (0.8804) \\
			
			MPLE.MS$ _o $(k)MS$ _o $(h)  &  38.980  & 0.6709  	&  -0.0713   	& -0.1212   & 0.2614  	&  0.3606  &  0.1235   \\
			& (0.005)  & (0.0029)  &  (0.0018)  	& (0.0000)  &           & (0.8858) &  (0.7685)   \\
			
			MPLE.MS$ _o $(k)MS$ _a $(h)  &  39.046  & 0.6164  	&  -0.0907   	& -0.2884  	& 0.2542 	&  0.3317  &  0.1155    \\
			& (0.006)  & (0.0034)  &  (0.0016)  	& (0.0036)  &           &  (0.9119) &  (0.8338)   \\
			
			MPLE.P$ _o $(k)MS$ _a $(h)   &  39.041  & 0.6343  	&  -0.0439   & -0.2388   & 0.2394 	&  0.3244  &  0.1105   \\	
			& (0.046)  & (0.0318)  &  (0.0106)  & (0.2141)  &          & (0.9181) &  (0.8710)  \\
			\noalign{\smallskip}\hline
	\end{tabular}}
\end{table}

\begin{figure}
	\begin{center}
		\includegraphics[scale=0.65]{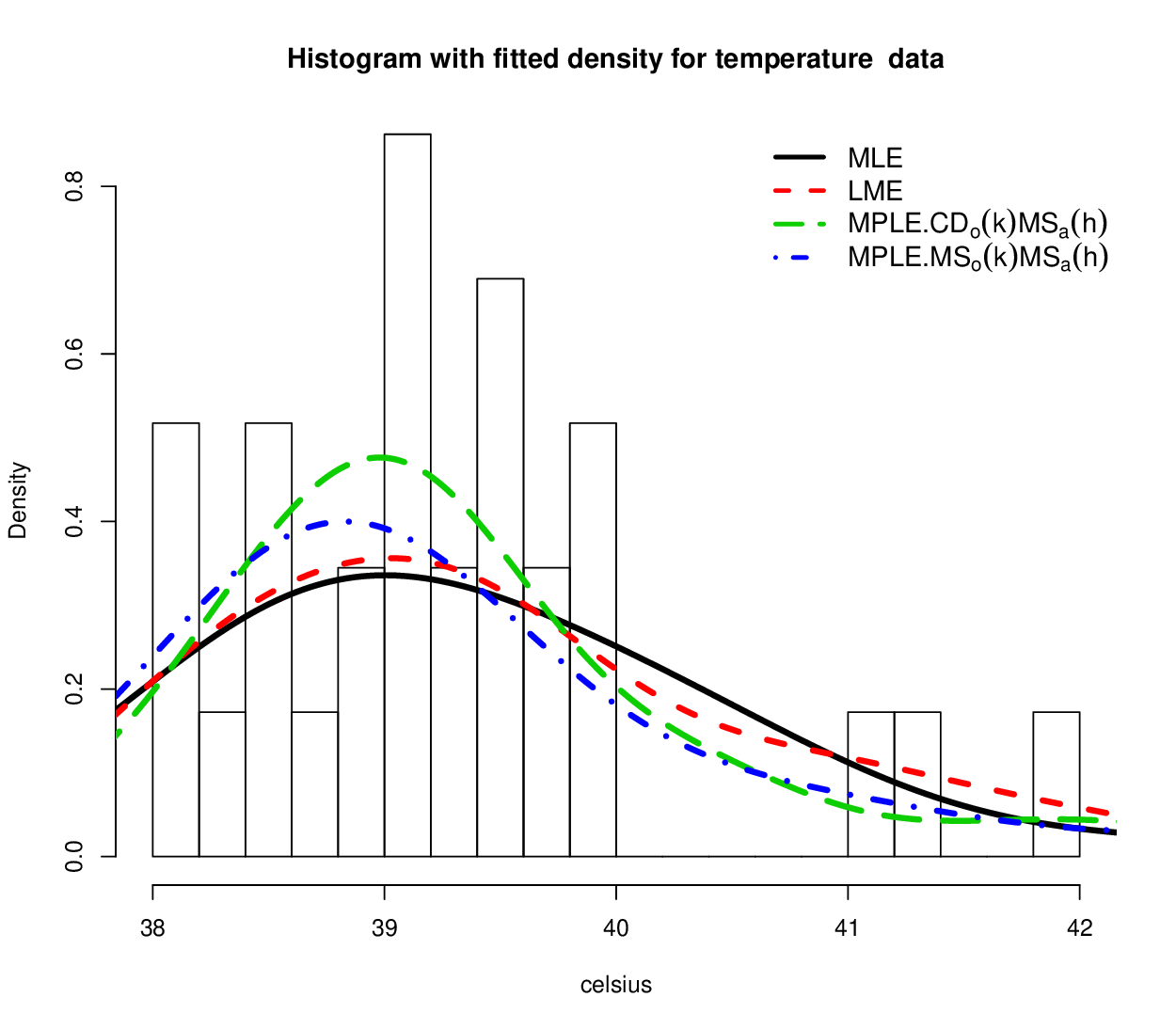}
		\caption{Histogram and the fitted densities of the four-parameter kappa distribution in which four different estimation methods are applied to the annual maximum daily temperatures (unit:$^{\circ} \text {C}$) in Surin.} \vspace{0.3cm}
		\label{tem_surin}
	\end{center}
\end{figure}

\begin{figure}
	\begin{center}
		\includegraphics[scale=0.65]{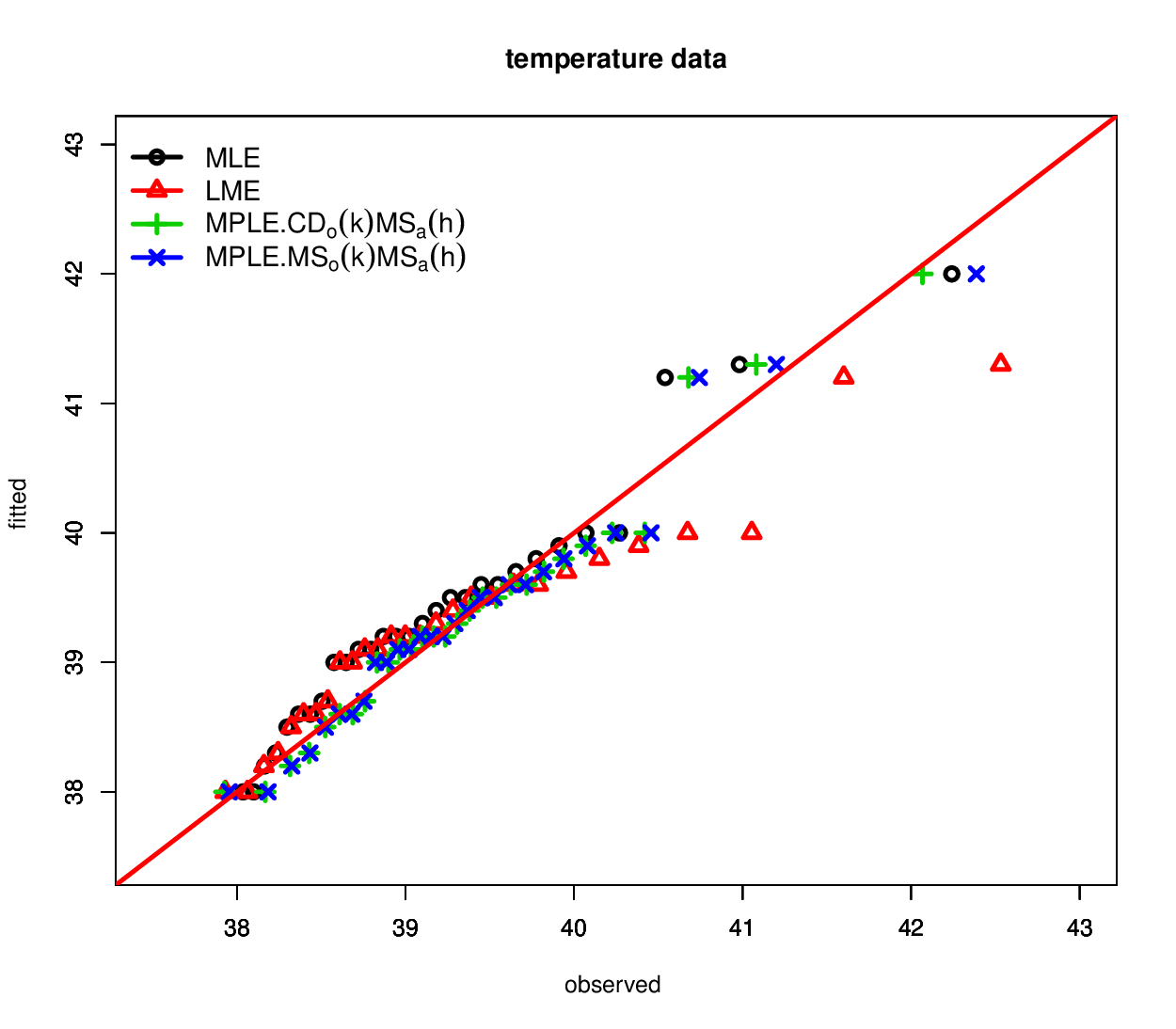}
		\caption{Same as Figure \ref{tem_surin} but quantile-quantile plot.} \vspace{0.3cm}
		\label{qqtemperature}
	\end{center}
\end{figure}

Table \ref{temperature} provides the estimates with standard errors and goodness-of-fit statistics with p-values. {The results indicate that 13 types of MPLEs show better performance than the MLE and LME in term MPAE, AD, and KS statistics. Specifically, the MPLE.CD$ _o $(k)MS$ _a $(h) is better fitted than any other estimation for this data set in term of estimates with standard error, KS statistics and MPAE.} The model diagnostic plots (histogram with density and qq-plot) from different estimation methods are presented in Figures \ref{tem_surin} and \ref{qqtemperature}. It suggests that MPLE.CD$_o $(k)MS$_a $(h) fits the temperature data reasonably well with K4D.

\begin{table}[!h]
	\caption{Same as Table \ref{temperature} but 20-year return level and 95$\%$ confidence interval.} \vspace{0.3cm}
	\label{20_year_temperature}
	\centering
	\scalebox{1}{
		\begin{tabular}{lcccc}
			\hline
			\multirow{2}{*}{\textbf{Methods} } & \multicolumn{4}{ c }{\textbf{20-year return level}} \\[1ex] \cline{2-5}
			
			& \hspace{0.3cm} Lower & \hspace{0.3cm} Point {(SE)}& \hspace{0.3cm} Upper  &  \hspace{0.3cm} Length    \\ [0.5ex]
			\noalign{\smallskip}\hline\noalign{\smallskip}
			MLE     &  41.09   & 41.18 (0.044)	&  41.27 & 0.18  \\
			LME    	&  41.73   & 42.47 (0.373)	&  43.22 & 1.50  \\
			MPLE.CD$ _o $(k)MS$ _a $(h)     &  40.90 	& 40.95 (0.027)	&  41.01 & 0.11  \\			
			MPLE.MS$ _o $(k)MS$ _o $(h)    	&  41.10 	& 41.12 (0.013)	&  41.15 & 0.05  \\
			MPLE.MS$ _o $(k)MS$ _a $(h)    	&  41.04 	& 41.09 (0.026)	&  41.14 & 0.10  \\
			MPLE.P$ _o $(k)MS$ _a $(h)    	&  40.92 	& 40.97 (0.024)	&  41.02 & 0.10 \\
			\noalign{\smallskip}\hline
	\end{tabular}}
\end{table}

\begin{figure}
	\begin{center}
		\includegraphics[scale=0.65]{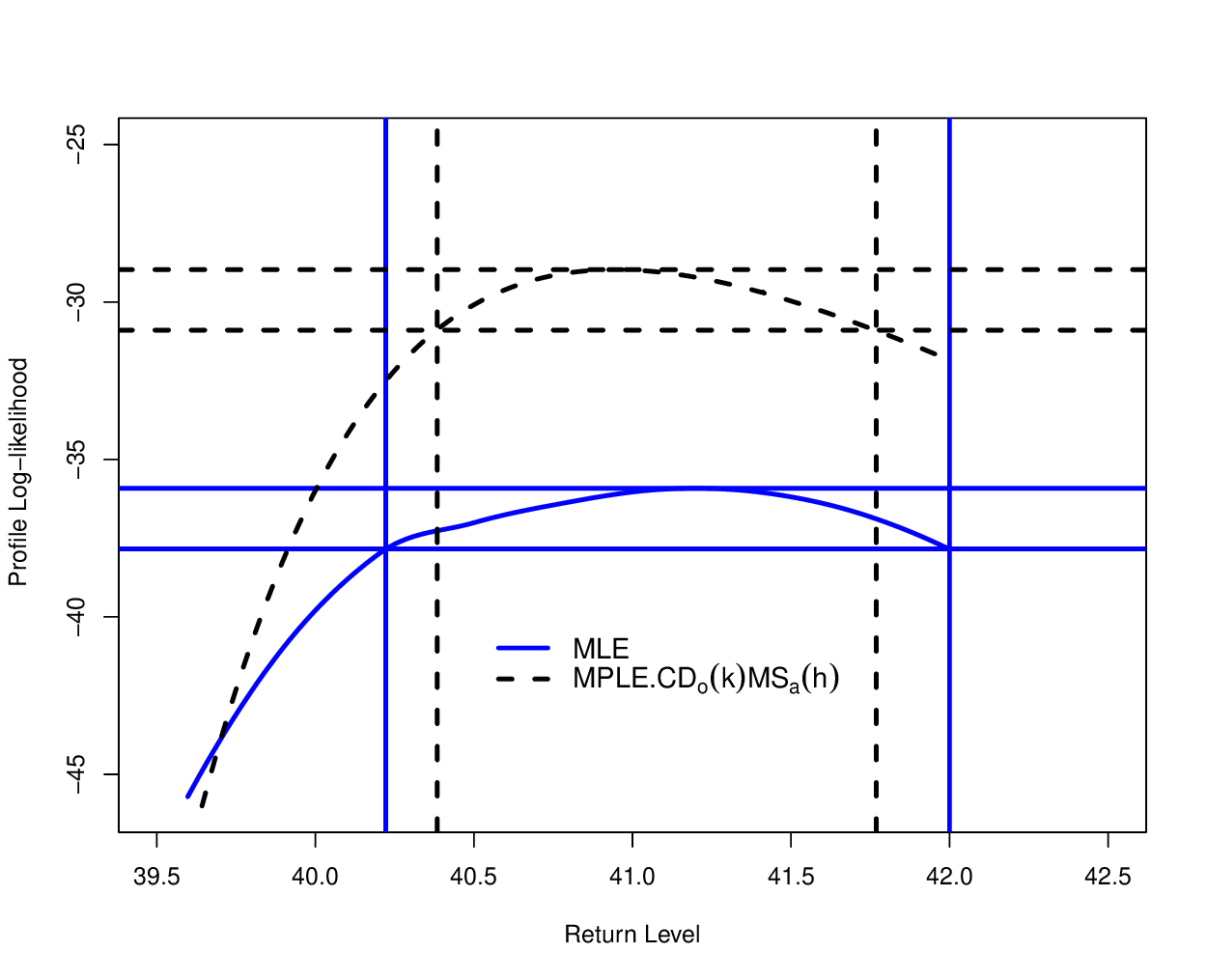}
		\caption{Profile likelihood, computed by theMPLE.CD$ _o $(k)MS$ _a $(h) method, for the 20-year return level of Surin's temperature data ($^{\circ} \text {C}$), which provides a 95$\%$ confidence interval of [40.45$^{\circ} \text {C}$, 41.89$^{\circ} \text {C}$].} \vspace{0.3cm}
		\label{profile_tem_surin}
	\end{center}
\end{figure}

Table \ref{20_year_temperature} shows a comparison at the 95\% confidence interval of the 20-year return level and standard error by MLE, LME, and MPLEs. To obtain the 95\% confidence interval for the 20-year return level, given by 41.03$^{\circ} \text {C}$, we employ the profile likelihood approach explicitly calculated using a reparameterization of K4D. Figure \ref{profile_tem_surin} shows that the 95\% confidence interval of the 20-year return level is [40.45$^{\circ} \text {C}$, 41.89$^{\circ} \text {C}$]. This interval is smaller than the one obtained from the MLE in Table \ref{temperature}. The details of reparameterization and the profile likelihood approach for K4D are provided in the Supplementary Material.

\section{Concluding remarks}\label{sec:7Conclusion}
This study used estimation methods based on maximum likelihood, L-moments, and maximum penalized likelihood to fit the K4D to data.
The K4D is a generalization of some common three-parameter distributions and, in particular, of the GEV distribution. It may be useful in fitting data when three parameters are not sufficient. The MLE is common and asymptotically optimal, but it has large variance for small sample sizes. The LME is usually considered better than the MLE for small samples, but it sometimes fails to not provide estimates due to convergence failure. This study therefore suggests using the MPLE method, which always provides estimates with smaller variance than the MLE. Modified versions of the existing three penalty functions are proposed for the second shape parameter ($h$) of K4D. A total of 18 penalty functions are tried.

Monte Carlo simulations are conducted to assess the effectiveness of the proposed estimation methods. As criteria to check their performance, we take the RBIAS and RRMSE on extreme upper quantiles. Based on this study, we suggest using the Coles and Dixon (CD) penalty for $k$ and Park (P) penalty for $h$ when the user does not have any prior information on the relevant data. When the user has some prior information that $k$ is negative, then using Martins and Stedinger’s (MS) penalty on $k$ and P (or MS or adjusted MS) penalty on $h$ is recommended. If the user has information that $k$ is positive, then CD (or P) penalty for $k$ and P (or adjusted CD) penalty on $h$ is recommended.


\section*{Acknowledgements}	
	The authors would like to thank Rajamangala University of Technology Isan, Thailand for granting financial support for this study. We would also like to thank the graduate school of Mahasarakham University for a research grant. Park's work was supported by the
	National Research Foundation of Korea (NRF) grant funded by the Korea
	government (MSIP) (No.2016R1A2B4014518).


\bibliographystyle{tfs}
\bibliography{bibPMLE}


\section*{Supplemental Material}
\begin{verbatim}
https://www.tandfonline.com/doi/suppl/10.1080/02664763.2021.1871592?scroll=top
\end{verbatim}

The following supporting information is available as part of the online article:

\noindent
\textbf{Figure S1 S2 and S3.}
{show some shapes of the probability density function (pdf) of K4D as the shape parameters ($ k,~h $) change, while $ \mu $ and $\sigma$ are set at 0 and 1, respectively.}

\noindent
\textbf{Figure S4.}
{The penalty function for various of $ p $ and $ q $ plotted against $ k $ }

\noindent
\textbf{Figure S5.}
{RBIAS of the 0.95 quantile estimators of K4D for sample size $ n = 30 $ with $ k = -0.4$ and $ -0.2 $.}

\textbf{Figure S6.}
{RBIAS of the 0.95 quantile estimators of K4D for sample size $ n = 30 $ with $ k = 0.2$ and $ 0.4 $.}

\noindent
\textbf{Figure S7.}
{RRMSE of the 0.95 quantile estimators of K4D for sample size $ n = 30 $ with $ k = -0.4$ and $ -0.2$.}

\noindent
\textbf{Figure S8.}
{RRMSE of the 0.95 quantile estimators of K4D for sample size $ n = 30 $ with $ k = 0.2$ and $ 0.4$.}

\textbf{Figure S9.}
{RBIAS of the 0.95 quantile estimators of K4D for sample size $ n = 50 $ with $ k = -0.4$ and $ -0.2 $.}

\textbf{Figure S10.}
{RBIAS of the 0.95 quantile estimators of K4D for sample size $ n = 50 $ with $ k = 0.2$ and $ 0.4 $.}

\noindent
\textbf{Figure S11.}
{RRMSE of the 0.95 quantile estimators of K4D for sample size $ n = 50 $ with $ k = -0.4$ and $ -0.2$.}

\noindent
\textbf{Figure S12.}
{RRMSE of the 0.95 quantile estimators of K4D for sample size $ n = 50 $ with $ k = 0.2$ and $ 0.4$.}

\noindent
\textbf{Table S1.}
{The RBIAS value of all estimation methods with shape paremeters $k=-0.2,h=-0.2.$}

\noindent
\textbf{Table S2.}
{The RRMSE value  of all estimation methods with shape paremeters $k=-0.2,h=-0.2.$}

\noindent
\textbf{Table S3.}
{The RBIAS value of all estimation methods with shape paremeters $k=-0.2,h = 0.2.$}

\noindent
\textbf{Table S4.}
{The RRMSE value  of all estimation methods with shape paremeters $k=-0.2,h=0.2.$}

\noindent
\textbf{Table S5.}
{The RBIAS value of all estimation methods with shape paremeters $k=0.4,h=-0.5.$}

\noindent
\textbf{Table S6.}
{The RRMSE value of all estimation methods with shape paremeters $k=0.4,h=-0.5.$}

\noindent
\textbf{Table S7.}
{The RBIAS value of all estimation methods with shape paremeters $k=0.4,h=0.5.$}

\noindent
\textbf{Table S8.}
{The RRMSE value of all estimation methods with shape paremeters $k=0.4,h=0.5.$}

\noindent
\textbf{Table S9.}
{Comparison of estimators and standard error of K4D fitted to temperature data with corresponding  MPAE, $KS$ and $AD.$}

\noindent
\textbf{Table S10.}
{Comparison of the 95\% confident interval of 20-year return level and standard error by MLE LM(GLD) and MPLEs.}

\end{document}